\documentstyle[aps,epsf,prl,multicol]{revtex}
\begin{document}
\draft
\title{  Localization in discontinuous quantum systems } 
\author{Fausto Borgonovi}
\address{
Dipartimento di Matematica e Fisica, Universit\`a Cattolica\\
via Trieste 17, 25121 Brescia, Italy  \\
I.N.F.N., Sezione di Pavia,
I.N.F.M., Unit\`a di Milano\\
}
\date{\today}
\maketitle 
 
\begin{abstract}
Classical and quantum properties of a discontinuous
perturbed twist map 
are investigated.
Different classical diffusive  
regimes, quasilinear and slow respectively,  are observed. 
The regime of slow classical diffusion gives rise
to two distinct quantal regimes, one marked by
dynamical localization, the other by quasi--integrable
localization due to classical Cantori.
In both cases the resulting quantum stationary
distributions are algebraically localized.
\end{abstract}
\pacs{PACS numbers: 05.45.+b }
 
\begin{multicols}{2}
A major feature of quantum dynamics of classically chaotic
systems is
the quantum suppression of classical chaotic excitations,
a phenomenon known as dynamical localization.
A prototype model,  both for classical chaos
and quantum dynamical localization is the Kicked
Rotator Model\cite{ccfi} (KRM) , whose  dynamics is 
described by the well known Chirikov standard map\cite{boris} (CSM).
This is a 2--d  continuous perturbed twist map, 
with a transition point, discriminating 
between bounded motion 
(prevalently regular on invariant KAM tori) and
unbounded and diffusive one (prevalently chaotic). 
Even though  transport properties of 2--d maps
are now quite well understood, analytical
results are only possible  for
particular maps, 
e.g. linear \cite{dana}.
In particular, the latter  are the simplest
discontinuous perturbed twist maps on the cylinder. 
For such  discontinuous  maps
the hypothesis of
KAM theorem are not satisfied and the
motion is typically unbounded
 even if it is possible  
to mark two different dynamical regimes 
(both diffusive).
Discontinuous maps also emerge from the study
of more concrete physical models, such as the 
motion of a particle colliding elastically
within  a two--dimensional bounded region
(billiard\cite{bcl}).
On the other side  very little is known about
the quantum dynamics of such discontinuous maps.
In particular, it is far from being obvious that  
 the relation between quantum 
localization and classical diffusion,
obtained for the KRM, holds
in this case too.

To answer the above questions, 
let us consider the following  
discontinuous 
map on the cylinder $[0,2\pi) \times [-\infty, \infty]$ 
\begin{equation}
\begin{array}{l} 
\bar{p} = p + k f(\theta) \nonumber \\
\bar{\theta} =  \theta + T \bar{p}  \quad {\rm mod}-2\pi\\
\label{dis}
\end{array}
\end{equation}
where $f(\theta) =  \sin (\theta)\ {\rm sgn} (\cos \theta ) $.
This function is a particularly
simple approximation of the stadium map\cite{bcl,bcr}. Moreover,
it is quite similar to the CSM (where $f(\theta)=\sin \theta$)
which has been widely investigated in the past.

Even if the following analysis has been put forward for
this specific function, it can be  generalized \cite{bcr}
to generic 
discontinuous, 
periodic and  bounded ($\vert f(\theta)\vert \leq 1 $)
 functions.
 This set of functions can be also
enlarged to continuous bounded functions with a 
discontinuous derivative. In this case the situation is
slightly complicated, since
 usually a critical value of the parameter
$K=kT$ appears (see \cite{bullet} for 
the piecewise linear  map)  such  that,
when $K < K_{cr}$,
the phase space is covered by invariant tori which do not permit
 unbounded motion along the cylinder:  only for
$K > K_{cr} $ the motion is diffusive.

\vspace{-2cm}
\begin{figure}
\hspace {1cm}
\epsfxsize 7cm
\epsfbox{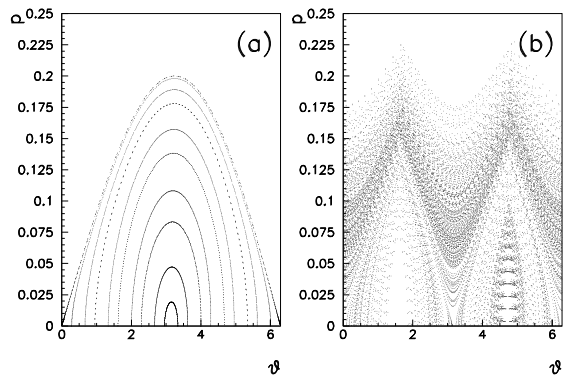}
\vspace{-2.5cm}
\narrowtext
\caption{Poincar\`e surface of section
for $k=0.01$, $T=1$. a) 10 different particles with
initial momentum $p=0.001$ and different phase $\theta$
have been
iterated  $n=10^3$ times for the
Chirikov standard map. (b) One particle
starting at the point $p_0 =0.011$
and $\theta_0 =3$
have been iterated $3\cdot 10^4$ times
using the discontinuous function $f(\theta)$.
}
\label{psos}
\end{figure}

The CSM is characterized by  
unbounded diffusive motion in the momentum $p$ for
$kT>1$ while, when $kT<1$, the motion is prevalently
regular with regions of stochasticity bounded by KAM invariant circles.
 On the other hand,
the classical properties of  map (\ref{dis}) are quite different.
Indeed, due to the discontinuities of $f(\theta)$ at 
$\theta = \pi/2, 3/2 \pi$, 
the hypothesis of KAM theorem are not satisfied
and, generally speaking, KAM tori does not exist, even  
for  
very  small $k$.
This means that one trajectory fills, in a dense way, any 
portion of the cylinder (phase space), for any $k \ne 0$.
Nevertheless the phase space is filled by Cantori\cite{robert}   
(remnants of KAM invariant tori) 
 which constitute partial barriers to the  motion\cite{percival}.
Due to the sticking of trajectories along these invariant structures
the diffusive motion is slowed down in close
analogy to the 
saw--tooth map case described in \cite{dana}.

An example of the classical
map dynamics is given in Fig.1. In the right 
picture (b),  
the Poincar\'e surface of section is shown for the discontinuous map
(\ref{dis}). A single 
 initial condition has been iterated $n=3\cdot 10^4$ times. As one can
see a single particle is free to wander in the whole phase 
space but the motion is far from being random.
Indeed, due to  sticking in the neighborhood of Cantori, 
the trajectory is almost regular on a finite time scale $\tau$.
Diffusive motion results from jumping among  
different stable varieties belonging to different Cantori.
As the iteration time, or the number of initial
particles, is increased, regular structures 
disappear  and
the surface of section appears to be covered uniformly.
For sake of comparison in the left picture (a) the same portion 
of phase space is shown for the CSM, with 
the same value of $k$. Here $10$ different trajectories have been
iterated $n=10^3$ times: each trajectory covers just one
torus. 
 
Despite the ``quasi'' regularity
 of 
the motion, numerical results  show that, 
 when $kT < 1$,  the dynamics is diffusive,
 for  $t > \tau$,
 along the cylinder axis  ($p-$coordinate)
with a diffusion rate $D$ given by

\begin{equation}
D = lim_{n\to\infty} { {\langle  p^2 (n) \rangle}\over {n}} 
= D_0\ k^{5/2} \sqrt{T}
\label{diff}
\end{equation}

where $n$ is the time measured in iterations of the map (\ref{dis})
and the average $\langle \ldots \rangle$ has been performed 
over an initial ensemble of particles with the same momentum $p$
and random phases $0< \theta < \pi $.
Also, in (\ref{diff}), $D_0 \simeq 0.4$ is a numerical constant (dependent from
the function $f(\theta)$) and the factor $\sqrt{T}$ has been added
for dimensional reasons. 
  
On the other side,
when $kT > 1$, the  
random phase approximation\cite{boris} can be applied and 
one finds diffusive motion along the $p-$direction
with a diffusion rate  $D \simeq D_{ql} = k^2/2$,
where $D_{ql}$ is the diffusion rate in the quasi--linear
approximation, namely assuming the phases $\theta$ to be completely
random uncorrelated variables.
Notice that, in the undercritical region $kT < 1$, the diffusion 
coefficient $D \simeq k^2 \sqrt{kT}$ is less than the quasilinear 
one $ D_{ql} \sim k^2$, due to the sticking of trajectories 
close to Cantori.
In Fig.2 the dependence of the diffusion rate $D$ is shown
as a function
of $k$
for $T=1$. The dashed and full lines
indicate respectively the quasilinear
 diffusion ($kT > 1 $) and the slow diffusion 
($kT < 1$). 

The apparently strange dependence of $D$ on $k$, in the ``slow''
diffusive
case $kT <1$ was  found in similar discontinuous maps,
e.g. the saw-tooth map\cite{dana} 
($f(\theta) =\theta/2\pi$),
 or the Stadium map\cite{bcl}.
In Ref.\cite{dana} a theoretical explanation of 
the exponent $5/2$ was given 
in terms of a Markovian model of transport  
based on the partition of phase space into resonances. 

\vspace {-1cm}
\begin{figure}
\hspace {0.3cm}
\epsfxsize 7cm
\epsfbox{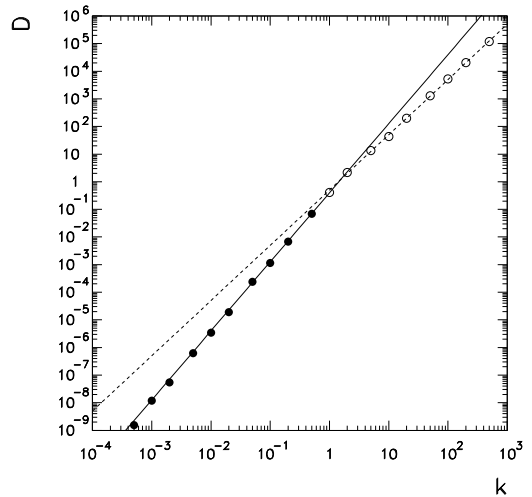}
\vspace{-1cm}
\narrowtext
\caption{
Diffusion rate for the discontinuous map 
as a function of $k$ and $T=1$.
Open and full circles indicate respectively the
``quasilinear'' and the ``slow'' diffusion.
Dashed line represents the quasilinear
approximation $D=D_{ql}=0.5 k^2$ which holds for $k>1$.
Full line is the best fit $D = 0.4 k^{5/2}$
obtained from full circles.
}
\label{difvk}
\end{figure}
 
Let us now consider  the quantized
version of  map (\ref{dis}).
According to a well--known procedure\cite{ccfi} the quantum
dynamics can be studied  by iterating 
the quantum evolution operator
over one period ${\cal U}_T$, starting from an initial
state $\psi_0 (\theta)$

\begin{equation}
\psi(T) = {\cal U}_T \psi_0 = e^{-i \hbar T{\hat n}^2/2} 
e^{-i k V(\theta) /\hbar }
\psi_0
\label{qmap}
\end{equation}

In (\ref{qmap}), as usual,  $\hat{n} = -i\hbar\partial/\partial \theta$
and $V(\theta) = \vert \cos \theta \vert$.
Quantum dynamics  depends on both parameters
$k/\hbar$ and $T\hbar$ separately. 
These parameters can be renormalized by letting $k/\hbar\to k$
and $T \hbar \to T$
(which is the same as to put $\hbar = 1$).
The semiclassical limit is then recovered by performing 
simultaneously the limits
 $k \to \infty$ and  $T\to 0$ keeping $kT=$ const.

The most studied example of quantization of twist maps like (\ref{dis})
is the KRM\cite{ccfi}, 
where $V(\theta) = \cos \theta$.
Nevertheless the regime $ k T < 1 $ , differently from
the case $ k T > 1 , \  k \gg  1 $, was not 
object of intense investigations.

At least numerically, one can observe two different regimes,
distinguished by the so--called Shuryak border $k=T$\cite{shuryak}.  
For $k>T$ the quantum steady state
is exponentially localized 
 over a number $l_\sigma \simeq \sqrt{k/T}$
of momentum states\cite{dima,felix}.
This number has been interpreted\cite{dima}, in a realistic way, 
as the number of quantized momentum 
states contained in the main classical
resonance (see Fig.1a) the size of which is $\sqrt{k/T}$\cite{boris}.
For $k<T$ the width of the
principal resonance is smaller than the distance among quantized 
momentum levels,
and no kind of semiclassical excitation process, based on the 
overlapping of resonances, is possible. 
In the following, the analysis will  then  be 
restricted to the case $k>T$ only. 
 
Before studying the  discontinuous case,
let us recall a few important facts  related to the
evolution operator (\ref{qmap}). 
Due to the discontinuity in the first derivative
of the potential $V(\theta)$,
the matrix elements of ${\cal U}_T$ in the 
 momentum basis decay according to  a power law  away from 
the principal diagonal:  $\vert {\cal U}_{n,n+s}\vert 
\simeq 1/s^2$. This case was  investigated\cite{fyodorov} 
for  Band Random matrices: it was found to be 
 typically characterized  by  power--law localized eigenstates
around their centers $n_0$, 
$\vert \phi_n \vert \simeq \vert n - n_0 \vert^{-2}$.

The following question is then important :
is it possible to connect 
quantum localization lengths and classical diffusion rates, 
as in the case of the KRM?
If so,
 what is the critical border necessary to start 
the classical-like diffusion process? 
Moreover, what is the r\^ole played by classical invariant
structures, such as  Cantori, in quantum dynamics? 

\vspace{-1cm}
\begin{figure}
\hspace {0.2cm}
\epsfxsize 7cm
\epsfbox{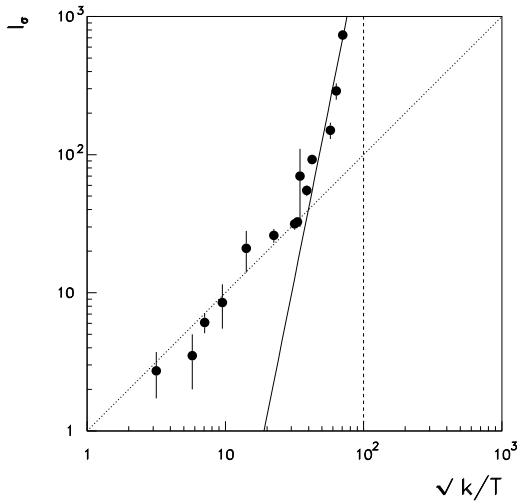}
\vspace{-1.5cm}
\narrowtext
\caption{
Localization length $l_\sigma$ as a function of
$ \protect\sqrt{k/T} $
for
fixed $T=0.01$.  Lines are the theoretical
predictions : dotted
($l_\sigma = \protect\sqrt{k/T}$),
full ($l_\sigma = D$).
Dashed line is the quasilinear border $k_{ql} T = 1$.
}
\label{loca}
\end{figure}

To answer the last question, let me remind the pioneering works
\cite{brown,geisel} where  quantum 
propagation of wave packets through 
the classical Cantori was first investigated. 
Other important results can be found in \cite{mackmei} where 
it was proposed that Cantori could  act,  
in Quantum Mechanics, as total barriers to the motion
if the flux exchanged through turnstiles is less than $\hbar$.
One can then reasonably assume that, in the deep quantum regime, 
the system will not be able to ``see'' the holes 
in the Cantori which behave as classical invariant tori. 

A more refined analysis requires the introduction of some kind of measure 
of the quantum distribution width. 
Since in this model localization 
is  presumably not exponential,
a unique  scale of localization
is not properly defined. For instance, while in the case 
of exponential localization the usual measures of localization,
e.g. inverse participation ratio, variance, entropy\cite{felix},
coincide, for power law localized distributions the dependence 
on the parameters can be different if different definitions
are adopted.
Then we choose the variance as a measure of the  distribution extension 
(degree of localization):

\begin{equation}
l_\sigma = [ \sum_n n^2 \vert \psi_n (t) \vert^2 ]^{1/2}
\label{ls}
\end{equation}
which 
has a proper 
semiclassical limit. Since this is, in general, an oscillatory 
function of the iteration time, a further average in time is necessary
in order to get time-independent results.

Numerical 
data are presented in Fig.3 where 
$l_\sigma$  has been plotted 
as a function of $\sqrt{k/T}$. 
Excluding oscillations, data follow, for $k<k_{cr}$,  the dotted line
$\sqrt{k/T}$, as for KRM.
Indeed, as one can see comparing Fig.1a and Fig.1b the principal 
resonance and the ``quasi'' principal resonance have roughly 
the same size.
This is a manifestation of the regularity imposed by  quantum
mechanics, or, in other words, of the discrete nature of
the quantum phase space.
This means that  classical discontinuous structures 
behave exactly as  continuous ones.

On the other side, since the classical discontinuous system 
is diffusive, the number of  occupied quantum states should 
increase on going into the semiclassical region.
Following known arguments for the dynamical localization,
one can expect the localization length to be 
given by 
the number of states inside a 
``quasi'' principal resonance ($\sqrt{k/T}$),
as soon as  it  equals numerically the classical diffusion coefficient.
In this way the critical value $k_{cr}$ can be obtained
by equating the following expressions:

\begin{equation}
l_\sigma \simeq \sqrt{k/T} \simeq D = D_0 k^{5/2} \sqrt{T}
\end{equation}

which gives the value $k_{cr} = 1/\sqrt{D_0 T}$.

It is important 
to notice that 
the ``quasi--integrable'' value $l_\sigma \simeq \sqrt{k/T}$
can survive
well above the threshold $k=1$ which is the value necessary
to start the classical--like diffusion process for the 
KRM when $kT >1$. 
Also, this kind of 
localization is not connected 
with  any classical--like
diffusive process, resulting instead from a quasi--periodic motion.
The absence of diffusive quantum motion, in the region 
$T<k < k_{cr}$ can be ascribed 
to a 
``dynamical '' diffusion rate $l_\sigma$  less than 
the size of the ``quasi''--principal resonance $l_\sigma \simeq \sqrt{k/T}$.
For instance, numerical
simulation indicates a localization length 
$l_\sigma \simeq 80\pm 10 \simeq \sqrt{k/T}$ 
for $k=10 \gg 1$, $T=1/1000$, while $D \simeq 0.4 k^{5/2} \sqrt{T} = 4$. 
In other words, in the region dominated by 
slow diffusion, the threshold for classical--like
diffusion  is $k > k_{cr} = 1/\sqrt{D_0 T}$ and not $k > 1$.

These theoretical predictions are confirmed by the numerical
data presented in Fig.3,  which closely follow the curve (full
line) 
$l_\sigma = D$ for $k_{cr} < k < k_{ql}$.
Here $k_{ql}$ stands for the border of validity of quasilinear 
diffusion :  $k_{ql}  = 1/T$.
This confirms and extends the validity of the dynamical localization
theory 
even in the presence of 
``slow'' diffusion and algebraic decay.
This last point can be directly observed in Fig.4a where 
the quantum steady state distribution $P(n)$, 
is shown together 
with the corresponding line $n^{-4}$.

\vspace{-2cm}
\begin{figure}
\hspace {0.2cm}
\epsfxsize 7cm
\epsfbox{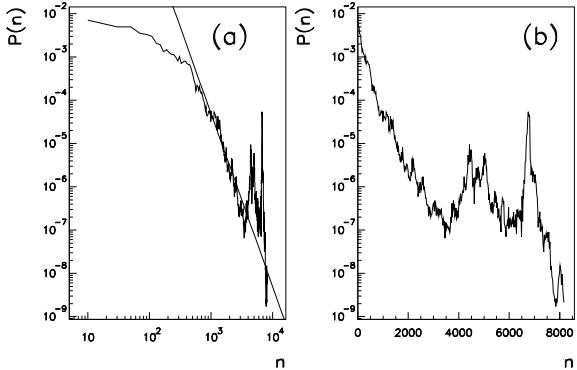}
\vspace{-2cm}
\narrowtext
\caption{
Quantum stationary distribution for $k=50$, and $T=0.01$.
The map has been iterated $10^6$ times. The
final distribution is obtained by averaging over the last $10^5$
kicks. The initial state is $\psi_n = \delta_{n,0}$.
(a) Log-log scale.  The line $n^{-4}$ has been drawn to guide
the eye. (b) Log scale.
}
\label{distr}
\end{figure}

The dynamical localization mechanism is not 
connected with this power law decay. Indeed the same algebraic
decay can be found for any $k$, in the region $k < k_{cr}$
as well  
for  $k > k_{ql}$ (at least in the tails of the 
distribution\cite{bcr}).
In more details, on semiclassically approaching 
 the  border $k_{ql}$, the 
quantum distribution shows big peaks 
of probability for high momentum values  
which indicate that 
new regions of the classical phase space
 are now quantally accessible. It is exactly the presence of such
peaks which causes the large increase of $l_{\sigma}$.
 The presence of  bumps of probability far from the initial state 
$n_0 = 0$ is shown in Fig.4b.

In conclusion,  a discontinuous map which is
a simple generalization of the Chirikov standard map
has been studied. Differently 
from the latter,
the dynamics is slowly diffusive even when the motion 
described by the CSM is prevalently regular.
In this region the quantum analysis reveals quite unexpected features.
Above the Shuryak border $k>T$, 
two different scaling laws for the 
localization length  are found.  The first, $l_\sigma \simeq \sqrt{k/T}$,
marked by the presence of classical Cantori acting as total barriers
to quantum motion, is a region of quantum integrability.
The second is a region characterized by 
 dynamical localization
($l_\sigma \simeq D$) thus indicating the existence of this phenomenon 
even in  case of slow diffusion. At the critical point $k_{cr}$,
separating these regimes,
  quantum dynamics starts
to follow the classical excitation process. 
Differently from the KRM, for which $k_{cr} \simeq 1$, one finds here
$k_{cr} \simeq 1/\sqrt{T}$. 
  
During the completion of this manuscript
I became aware of another related work\cite{prange}
where a regime of quantum integrability 
is found, for the Stadium billiard, in the region delimited by  
the inequalities $E\epsilon > 1$ and
$\sqrt{E}\epsilon^2 <1$, where $E$ is the energy 
of the particle and $\epsilon\ll 1$ 
is the
ratio between the straight line and the circle radius\cite{bcl}.
The billiard
dynamics is well described\cite{bcr} in terms of the map (\ref{dis})
via the substitutions $k = 2\epsilon\sqrt{E}$, $T=\sqrt{2/E}$.
It is then easy to verify that the 
quantum--integrable regime found in Ref.\cite{prange}
$E^{-1} < \epsilon < E^{-1/4}$, coincides with 
 the regime dominated by classical Cantori $ T < k < 1/\sqrt{T}$.
 
This may be a first indication that not only the classical, 
but also the quantum dynamics of the Stadium, 
can be  described in terms of maps:   
this will be the subject of a future work\cite{bcr}.

The author is thankful to G.Casati, I.Guarneri and D.L.Shepelyansky
for useful discussions.

\end{multicols}
\end{document}